# Correlations of Spectroscopic and Dielectric Properties of Hafnia-Zirconia Nanoparticles

Yuriy O. Zagorodniy[1], Eugene A. Eliseev[1*], Petr Jiricek[2], Jana Houdkova[2], Lesya Demchenko[3,4], Oksana V. Leshchenko[1], Victor N. Pavlikov[1], Ihor V. Pleskach[1], Anna O. Diachenko[5], Michail D. Volnyanskii[5], Oleksandr S. Pylypchuk[6], Myroslav V. Karpets[4], Mikhail P. Trubitsyn[2,5†], and Anna N. Morozovska[6‡]

[1] Frantsevich Institute for Problems in Materials Science of the National Academy of Sciences of Ukraine, 3, str. Omeliana Pritsaka, 03142 Kyiv, Ukraine

[2]Institute of Physics of the Czech Academy of Sciences, Na Slovance 1999/2, 18200 Prague 8, Czech Republic

[3] Stockholm University, Department of Chemistry, Sweden

[4] Ye. O. Paton Institute of Materials Science and Welding, National Technical University of Ukraine "Igor Sikorsky Kyiv Polytechnic Institute", 37, Beresteisky Avenue, Kyiv, Ukraine, 03056

[5] Oles Honchar Dnipro National University, 72, Nauky Avenue, 49000 Dnipro, Ukraine

[6] Institute of Physics of the National Academy of Sciences of Ukraine, 46, Nauky Avenue, 03028 Kyiv, Ukraine

## Abstract

In this work we analyze correlations of X-ray photoelectron and diffraction spectra, and dielectric properties of hafnia-zirconia nanoparticles (the chemical compositions $Hf_{0.4}Zr_{0.6}O_2$ and $Hf_{0.6}Zr_{0.4}O_2$, and the average size of 7.5 nm) prepared by the solid-state organo-nitrate synthesis and annealed in air. The phase state of the nanoparticles, determined by the X-ray diffraction spectroscopy, is the coexistence of the nonpolar monoclinic (42 – 76 mass %) and orthorhombic (56 – 24 mass %) phases. Concentration of the oxygen vacancies was estimated from the X-ray photoelectron spectroscopy. The increase in the intensity of the dielectric permittivity maximum observed near 350 – 450 K in a PVDF matrix with embedded $Hf_{0.4}Zr_{0.6}O_2$ nanoparticles can be related with an increase in oxygen vacancy concentration. Theoretical calculations, based on Landau-Ginzburg-Devonshire theory, explain the increase of the dielectric permittivity in $Hf_{0.4}Zr_{0.6}O_2$ nanoparticles compared to $Hf_{0.6}Zr_{0.4}O_2$ nanoparticles.

[*]Corresponding author, e-mail: eugene.a.eliseev@gmail.com

[†] Corresponding author, e-mail: trubitsyn_m@ua.fm

[‡]Corresponding author, e-mail anna.n.morozovska@gmail.com

## I. INTRODUCTION

The discovery of ferroelectricity in thin films of binary oxides such as $HfO_2$ and $ZrO_2$ [1, 2, 3] made these materials as highly promising candidates for non-volatile memory devices and integrated layered capacitors [4, 5, 6]. Unlike perovskite ferroelectrics, which suffer from poor compatibility with silicon technologies due to substantial lattice mismatches and the formation of interfacial dead layers that degrade their ferroelectric response, $(Hf,Zr)O_2$ thin films are free from these deficiencies and thus are successfully used as gate dielectrics in field-effect transistors. Their chemical stability and full compatibility with complementary metal-oxide-semiconductor (CMOS) processes make them particularly attractive for modern microelectronic applications.

The ferroelectric (FE) properties of $(Hf,Zr)O_2$ thin films are associated with the formation of polar orthorhombic phase (o-phase) with the non-centrosymmetric space group $Pca2_1$. In contrast, bulk $HfO_2$ and $ZrO_2$, are nonpolar centrosymmetric materials. At room temperature and normal pressure, both $HfO_2$ and $ZrO_2$, which are structurally equivalent, stabilize in the monoclinic phase (m-phase) with the space group $P2_1/c$. With increasing temperature, this phase undergoes a transformation to the tetragonal phase (t-phase) with the space group $P4_2/nmc$, followed by a transition to the cubic phase (c-phase) with the space group *Fm3m* [7, 8]. Under high-pressures, two additional nonpolar o-phases (o-I and o-II) with the space groups *Pbca* and *Pnma* can be stable [9, 10].

The formation of the FE phase in $(Hf,Zr)O_2$ thin films can be attributed to several factors, including elastic strains arising from deposition on substrates with different lattice parameters [11, 12, 13], changes in the electronic structure due to doping, lattice distortions caused by the formation of oxygen vacancies [14, 15], size effects [16], as well as due to the interplay of all these mechanisms.

It was later shown experimentally that oxygen-deficient $Hf_xZr_{1-x}O_{2-y}$ nanoparticles, which contain a large fraction of o-phases due to the annealing in $CO$+$CO_2$ ambient [17, 18, 19], can exhibit ferroelectric-like properties, such as a colossal dielectric response over a wide frequency range [20, 21], as well as demonstrate resistive switching and pronounced charge accumulation [22].

Also, it was shown theoretically [23, 24, 25] that the influence of vacancies on the stabilization of the FE o-phase arises from the formation of elastic dipoles around point defects, which leads to the local breaking of the inversion symmetry. The density functional theory (DFT) calculations, performed in Ref. [26], confirmed the influence of oxygen vacancies on the phase stability of $HfO_2$ and revealed that the FE o-phase becomes energetically more favorable than the m-phase, although it remains less favorable than the tetragonal and cubic phases. However, the physical mechanisms governing the stabilization of the FE phase are not sufficiently studied.

In this work we analyze correlations of X-ray photoelectron and diffraction spectra, and dielectric properties of hafnia-zirconia nanoparticles (the chemical compositions $Hf_{0.4}Zr_{0.6}O_2$ and $Hf_{0.6}Zr_{0.4}O_2$, and the average size of 7.5 nm) prepared by the solid-state organo-nitrate synthesis and annealed in air. The phase state of the nanoparticles, determined by the X-ray diffraction spectroscopy, is the coexistence of the nonpolar monoclinic (42 – 76 mass %) and orthorhombic (56 – 24 mass %) phases. Concentration of the oxygen vacancies was estimated from the X-ray photoelectron spectroscopy. The increase in the intensity of the dielectric permittivity maximum observed near 350 – 450 K in a PVDF matrix with embedded $Hf_{0.4}Zr_{0.6}O_2$ nanoparticles can be related with an increase in oxygen vacancy concentration. Theoretical calculations, based on Landau-Ginzburg-Devonshire theory, were performed to explain the difference of the dielectric permittivity of $Hf_{0.4}Zr_{0.6}O_2$ and $Hf_{0.6}Zr_{0.4}O_2$ nanoparticles.

## 2. Samples Preparation and Experimental Techniques

To study how oxygen vacancies affect the emergence of possibly ferroelectric o-phase, we prepared $(Hf,Zr)O_2$ nanoparticles by the solid-state organonitrate synthesis. The process employed aqueous mixtures of zirconium and hafnium nitrate salts ($ZrO(NO_3)_2{\cdot}2H_2O$ and $Hf(NO_3)_2{\cdot}2H_2O$), dissolved in distilled water at concentrations not exceeding 5 – 10% (for methodological details, see Refs. [17, 18]). Nanoparticles of hafnia-zirconia which chemical composition $Hf_{0.4}Zr_{0.6}O_2$ (the sample S1) and $Hf_{0.6}Zr_{0.4}O_2$ (the sample S2) were annealed in air at 700 °C for 6 hours.

X-ray studies of the samples S1 and S2 were performed using an Ultima IV X-ray diffractometer (Rigaku, Japan) with monochromatic Cu Kα radiation, and with the scanning step of 0.04°, and the exposure time per point 2 s. The X-ray diffraction data were analyzed by the Rietveld method using the FullProf software package [27], and the sizes of the coherent scattering region (CSR) were calculated.

Composite materials were prepared using the hafnia-zirconia nanoparticles from the samples S1 and S2. The nanoparticles, which occupied approximately 13 vol. % of the total composite volume, were embedded in the PVDF polymer matrix and then the composite plates compressed (~100 MPa). Silver electrodes were deposited on the faces of the samples. The electrical properties of the $Hf_{0.4}Zr_{0.6}O_2$ – PVDF and $Hf_{0.6}Zr_{0.4}O_2$ – PVDF composites were measured in an AC field ($f \approx 10^2 - 10^6$ Hz) using a Keysight E4980AL LCR meter from the room temperature up to ~440 K, above which the PVDF-based composites significantly softened due to heating, making reliable measurements impossible.

The X-ray photoelectron spectra (XPS) were measured by the AXIS Supra photoelectron spectrometer using monochromatized Al Kα radiation (1486.6 eV, 300 W, area analyzed – $0.7{\times}0.3$ mm$^2$). A source of gas cluster ions was used to remove contaminants from the surface. To minimize the damage of the sample surface due to interaction with high-energy ions, the ion source operated in the cluster mode (the energy of the primary cluster ion was 5 keV, and the number of argon atoms in the cluster was 2000). This mode of the ion source reduces significantly the surface damage of the samples. The irradiation time with primary cluster ions was 5 minutes. Photoelectron lines were recorded before and after the etching process. The main peak of the C 1s spectrum, corresponding to carbon contamination (C–C) was set at 284.5 eV.

## 3. Experimental results and discussion

The transmission electron microscopy (TEM) images show that the $(Hf,Zr)O_2$ nanoparticles have individual sizes around 7.5 nm and high crystallinity degree (see **Fig. 1**, top row). The electron dispersion spectra (EDS) analysis showed that the Hf, Zr, and O elements are uniformly distributed in the $(Hf,Zr)O_2$ nanoparticles, forming a solid solution (see **Fig. 1**, bottom row). According to the scanning electron microscopy analysis (see Ref. [19] for details), the samples S1 – S2 are homogenous and consist of the clusters with the average size of about 65 nm.

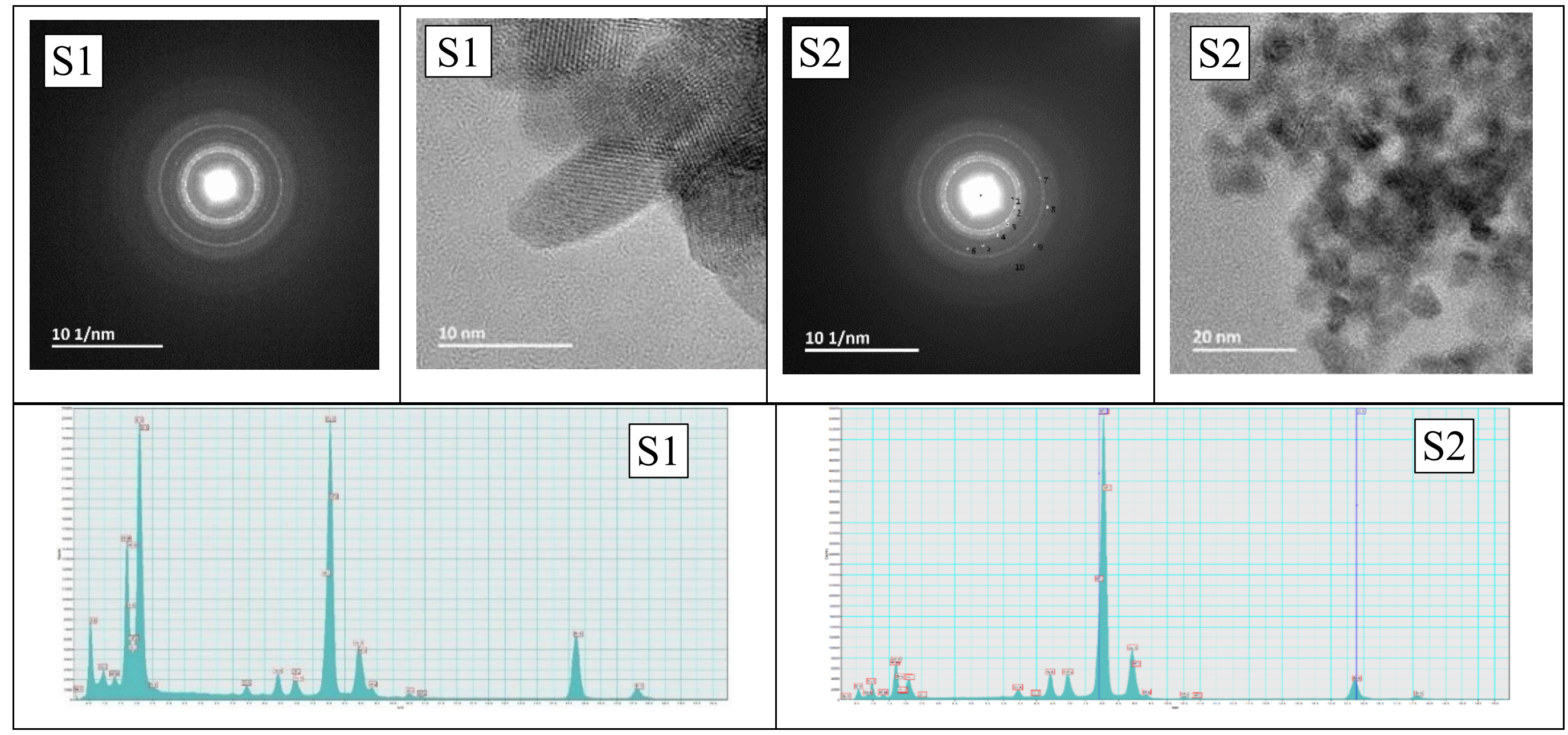

**Figure 1.** TEM images (top row) and EDS spectra (bottom row) of the samples S1 and S2.

The X-ray diffraction (XRD) spectra of the samples S1 and S2 are presented in **Fig. 2(a)** and **2(b)**, respectively. While the spectrum of the monoclinic phase differs significantly from other possible phases, the spectra of the tetragonal and three orthorhombic phases are very difficult to distinguish from each other using X-ray diffraction spectra. The three orthorhombic phases are even more difficult to distinguish, as they have almost identical cation arrangement and differ only in the positions of the oxygen atoms [28]. Based on the full-profile analysis of the XRD spectra, the nanopowders predominantly display a coexistence of monoclinic (m) and orthorhombic-like (o) phases. Phase composition (in mass %) and CSR size (in nm), determined from the XRD, are listed in the last column of **Table 1**.

Direct measurements of the nanopowder dielectric permittivity cannot be reliable due to the possible redistribution of the nanoparticles during the measurement, and therefore results depend on the pressure applied to the powder cell [19]. Therefore, we measured and analyzed the effective dielectric permittivity of the composites consisting of hafnia-zirconia nanoparticles homogeneously distributed and fixed in a PVDF matrix. The studied nanocomposites $Hf_{0.4}Zr_{0.6}O_2$ − PVDF and $Hf_{0.6}Zr_{0.4}O_2$ − PVDF contain about 13 vol. % of the hafnia-zirconia nanoparticles annealed in air, which were taken from the nanopowder samples S1 and S2.

Temperature dependences of the effective dielectric permittivity $\varepsilon_{eff}^*$ of the nanocomposites $Hf_{0.4}Zr_{0.6}O_2$ – PVDF and $Hf_{0.6}Zr_{0.4}O_2$ – PVDF, measured at the frequencies from ~250 Hz to 1 MHz, are shown in **Fig. 2(c)** and **2(d)**, respectively. The permittivity $\varepsilon_{eff}^*$ of the $Hf_{0.4}Zr_{0.6}O_2$ − PVDF plates (sample S1+PVDF) is significantly larger (~ 3 times) than that of the $Hf_{0.6}Zr_{0.4}O_2$ – PVDF plates (sample S2+PVDF). At low frequencies, the temperature dependence of $\varepsilon_{eff}^*$ has several smeared features at approximately 350 K (local minima) and 430 K (maxima). These features decrease and vanish as the frequency increases from 250 Hz to 1 MHz. The magnitude of $\varepsilon_{eff}$ decreases monotonically with an increase in frequency.

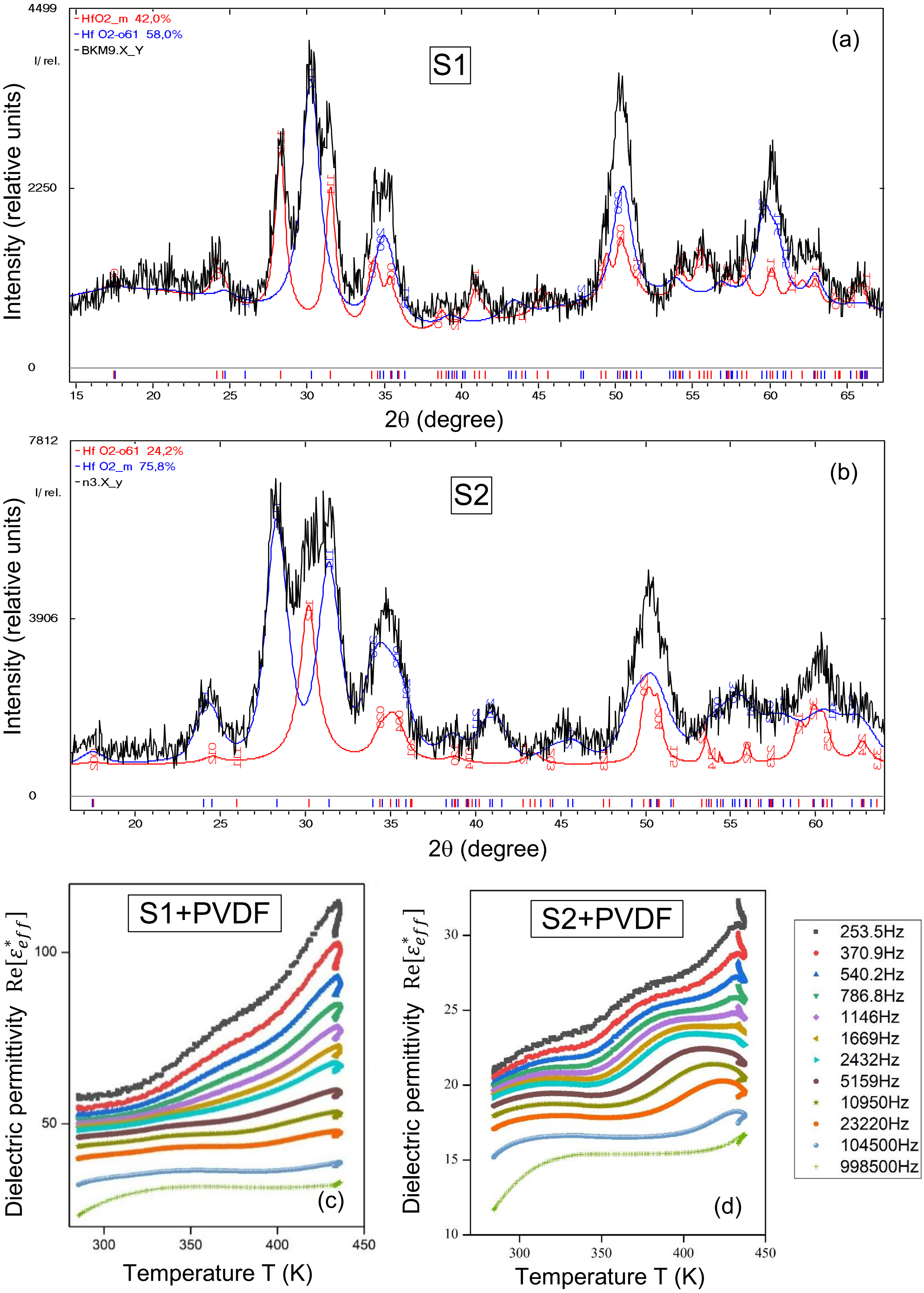

**Figure 2**. XRD spectra of the powder samples S1 (nanoparticles) **(a)** and S2 ($Hf_{0.6}Zr_{0.4}O_2$ nanoparticles) **(b),** which contain $Hf_{0.4}Zr_{0.6}O_2$ or $Hf_{0.6}Zr_{0.4}O_2$ nanoparticles annealed in air. Temperature dependences of the real part of the effective dielectric permittivity $\varepsilon^*_{eff}$ of the pressed $Hf_{0.4}Zr_{0.6}O_2$ – PVDF **(c)** and $Hf_{0.4}Zr_{0.6}O_2$ – PVDF **(d)** nanocomposites measured in the frequency range from 250 Hz to 1 MHz (see legend).

To separate contribution of the possible defects and surface contamination, the Hf 4f, Zr 3d and O 1s XPS spectra were obtained for as-prepared samples S1 and S2 and the samples after "soft" etching their surface for 300 seconds with argon ions. To obtain more details about the formation of chemical bonds by oxygen atoms and, possibly, to track the formation of oxygen vacancies, a detailed study of the high resolution O1s XPS spectra was performed (see **Fig. 3**). The O1s XPS spectra of the as-prepared samples (see **Fig. 3(a, c)**) shows a highly asymmetric peak, which can be decomposed into three Gaussian lines with binding energies of about 530, 531.5 and 532.2 – 533.8 eV. They are designated as I, II, III in **Fig. 3**. The change in the number of defects presumably induced in the samples (depending on the synthesis conditions) is accompanied by a change in the asymmetry of the spectral line, which is manifested in a greater relative intensity of the peaks II and III. The binding energies of 1s electrons for all lines obtained from the spectra decomposition on Gaussians, as well as the contribution of individual peaks, are presented in **Table 1**, which also provides the contribution of the orthorhombic phase in the studied samples, determined from the X-ray diffraction spectra.

The most intensive peak I in these compounds corresponds to the line with the binding energy of about 530 eV and can be assigned to lattice oxygen, which has similar values in the $HfO_2$ and $ZrO_2$ [29, 30], where the Hf and Zr ions have an oxidation state of +4 and oxygen vacancies or other defects are absent. Peaks II and III, by their position, can belong to hydroxyl groups, which may be present in the samples S1 – S2 due to the synthesis conditions.

The relative contribution of lines I–III changes significantly in the XPS spectra of the etched samples (see **Fig. 3** (**b, d**). For all samples, a sharp decrease in the intensity of the peak II is observed (see **Table 1**), which allows us to conclude that this peak includes signals from the contaminants of the surface as well as from weakly bounded molecules on the sample surface. Meanwhile, the relative intensity of the peak III with higher binding energy increases. This result can be explained by the role of oxygen vacancies, which act as active surface centers and promote the absorption of water molecules with subsequent dissociation some of them to the OH groups depending on the number of electrons captured by the vacancy. This also implies that water and OH groups adsorbed by the sites of oxygen vacancies are bound to the surface more strongly.

The significantly different content of the o-phase in the samples of $Hf_{0.4}Zr_{0.6}O_2$ and $Hf_{0.6}Zr_{0.4}O_2$ annealed in air with approximately the same number of defects, as well as the significant decrease in relative amount of the o-phase with the increase in zirconium, may indicate that the formation of the o-phase under the influence of defects is more favorable for a higher zirconium content (see **Table 1**). Analysis of the experimental XRD, XPS and dielectric spectra therefore indicates that the $Hf_{0.4}Zr_{0.6}O_2$ nanoparticles are stabilized predominantly in the o-phase, which is not the case for the $Hf_{0.6}Zr_{0.4}O_2$ nanoparticles.

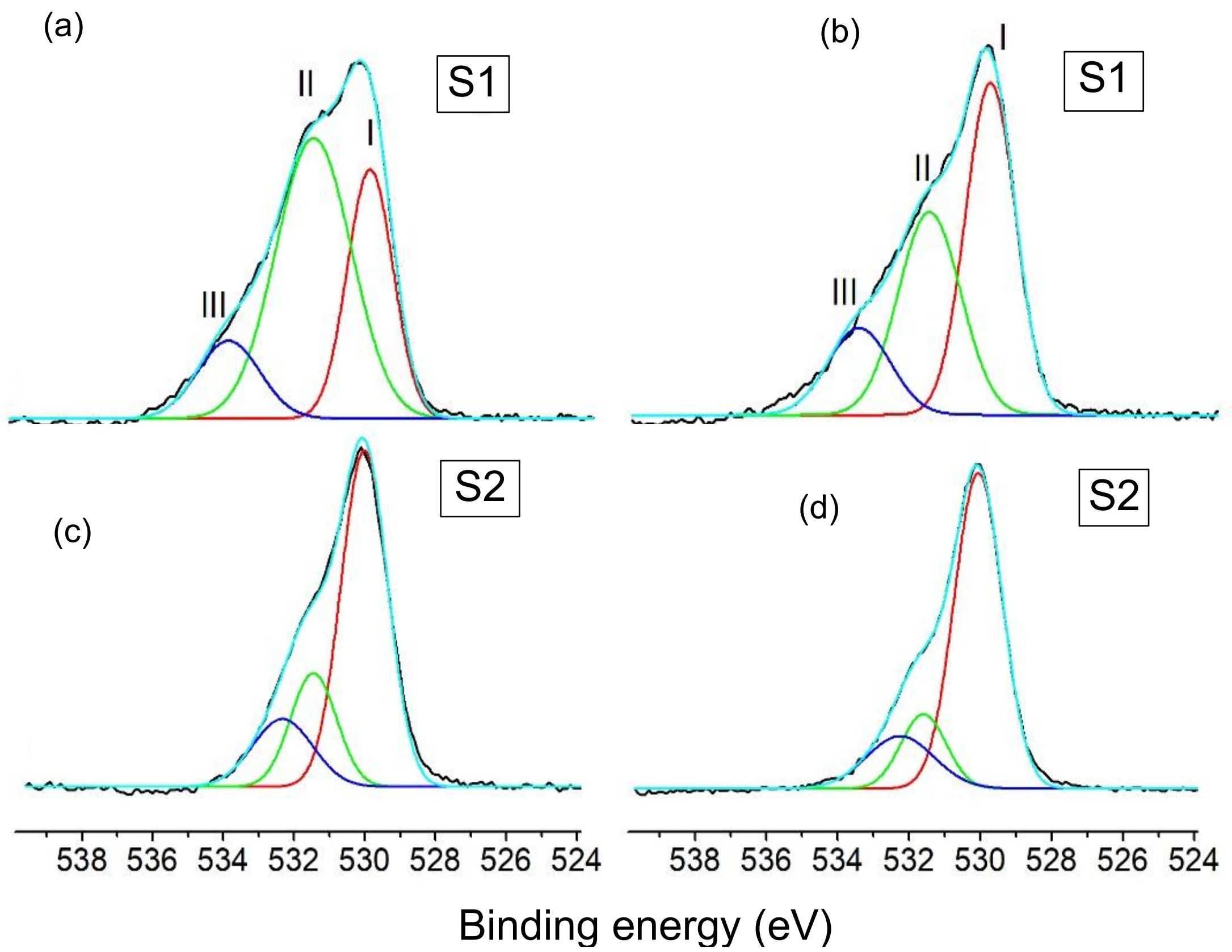

**Figure 3.** O1s XPS spectra of the studied as-prepared samples S1 and S2 (the left column **(a, c)**) and the samples after etching their surface during 300s (the right column **(b, d)**). Black curves are experimentally measured spectra, cyan curves are their fitting by three Gaussians shown by the red, green and blue colors. The preparation conditions of the samples S1 and S2 are described in **Table 1**.

**Table 1.** Preparation conditions of the samples S1 and S2, the binding energy and the peak area obtained in the result of the Gaussian decomposition of O1s XPS spectra, and the fraction of the o-phase in the samples

| Sample | | | Binding Energy, eV | | | Peak Area, % | | | Phase composition (mass %) and CSR size (nm) determined from XRD |
|---|---|---|---|---|---|---|---|---|---|
| | | | I | II | III | I | II | III | |
| S1 | $Hf_{0.4}Zr_{0.6}O_2$ annealed in air | As prepared | 530 | 531.5 | 533.8 | 30 | 57 | 14 | m-phase 42,02 mass %, CSR=11 nm. o-phase 57,98 mass %, CSR=8 nm |
| | | After etching | 530 | 531.5 | 532.9 | 56 | 14 | 30 | |
| S2 | $Hf_{0.6}Zr_{0.4}O_2$ annealed in air | As prepared | 530 | 531.5 | 532.3 | 63 | 21 | 16 | m-phase 75,85 mass %, CSR=12 nm o-phase 24,15 mass %, CSR=10 nm |
| | | After etching | 530 | 531.6 | 532.2 | 69 | 15 | 16 | |

## 3. Theoretical modelling

To describe the phase diagrams of the of $Hf_{0.4}Zr_{0.6}O_2$ and $Hf_{0.6}Zr_{0.4}O_2$ nanoparticles, we use the Landau-Ginzburg-Devonshire (LGD) approach proposed in Refs. [16, 19, 25]. The form of the free energy functional is based on Delodovici et al. works [31, 32], and Jung and Birol works [33, 34], where the principal role of the trilinear coupling between the polar, antipolar and nonpolar modes has been established. The LGD free energy of the core of $Hf_{0.4}Zr_{0.6}O_2$ or $Hf_{0.6}Zr_{0.4}O_2$ nanoparticles, corresponding to the ferroelectric o-phase, consists of the bulk energy density $f_{bulk}$, electric energy $f_{el}$ and the surface energy $F_s$ [16, 19, 25]:

$$F_{o-phase} = \int (f_{bulk} + f_{el}) dV + F_s, \quad f_{bulk} = f_{bq} + f_{tr} + f_{est} + f_{grad}. \tag{1}$$

The bulk energy density $f_{bulk}$ is the sum of the biquadratic energy $f_{bq}$ and trilinear coupling energy $f_{tr}$ of the polar, antipolar and nonpolar order parameters, elastic and striction energy contributions $f_{est}$, and the gradient energy of the order parameters $f_{grad}$. The energy $f_{bq}$ is an expansion over the even powers and $f_{tr}$ is an expansion over the odd powers of the dimensionless amplitudes $Q_{\Gamma 3}$, $Q_{Y2}$ and $Q_{Y4}$ of the polar phonon mode $\Gamma_{3-}$, nonpolar phonon mode $Y_{2+}$ and antipolar phonon mode $Y_{4-}$. The energies $f_{bq}$, $f_{tr}$, $f_{est}$, $f_{grad}$ and $f_{el}$ are listed in Refs. [16, 19, 25]. Corresponding material parameters are listed in Ref. [16].

The polarization $P_3$ is proportional to the amplitude $Q_{\Gamma 3}$ of the $\Gamma_{3-}$ mode [32, 33] and the dielectric permittivity $\varepsilon_{33}$ is proportional to its derivative over external field $E_3$:

$$P_3 = \frac{Z_B^* d}{V_{f.u.}} Q_{\Gamma 3} \approx P_0 Q_{\Gamma 3}, \qquad \varepsilon_{33} = \frac{\partial P_3}{\partial E_3}. \tag{2}$$

Here $P_0 \approx \frac{Z_B^* d}{V_{f.u.}}$ is the polarization amplitude of $Hf_{0.4}Zr_{0.6}O_2$ or $Hf_{0.6}Zr_{0.4}O_2$ material at room and lower temperatures [32].

The ferroelectric state of the core can be stable when its free energy $F_{o-phase}$ is smaller than the energy of the monoclinic phase, $F_m = \int f_m dV$, where $f_m$ is the energy density of the monoclinic phase in a bulk of $Hf_{0.4}Zr_{0.6}O_2$ and $Hf_{0.6}Zr_{0.4}O_2$. The increase of the dielectric permittivity in $Hf_{0.4}Zr_{0.6}O_2$ nanoparticles compared to $Hf_{0.6}Zr_{0.4}O_2$ nanoparticles, can be explained by the significant contribution of the nonpolar and antipolar orders to the permittivity due to the trilinear coupling in the free energy (1).

## IV. CONCLUSIONS

We analyze correlations of X-ray photoelectron and diffraction spectra, and dielectric properties of hafnia-zirconia nanoparticles with the chemical compositions $Hf_{0.4}Zr_{0.6}O_2$ and $Hf_{0.6}Zr_{0.4}O_2$, and the average size of 7.5 nm, prepared by the solid-state organo-nitrate synthesis and annealed in air.

The phase state of the nanoparticles, determined by the X-ray diffraction spectroscopy, is the coexistence of the nonpolar monoclinic (42 – 76 mass %) and orthorhombic (56 – 24 mass %) phases.

Concentration of the oxygen vacancies was estimated from the X-ray photoelectron spectroscopy. The increase in the intensity of the dielectric permittivity maximum observed near 350 – 450 K in a PVDF matrix with embedded $Hf_{0.4}Zr_{0.6}O_2$ nanoparticles can be related with an increase in oxygen vacancy concentration.

Theoretical calculations, based on Landau-Ginzburg-Devonshire theory, were performed. According to them, the increase of the dielectric permittivity in $Hf_{0.4}Zr_{0.6}O_2$ nanoparticles compared to $Hf_{0.6}Zr_{0.4}O_2$ nanoparticles, can be explained by the significant contribution of the nonpolar and antipolar orders to the permittivity due to the trilinear coupling in the Landau-Ginzburg-Devonshire free energy.

## Authors' contribution

Y.O.Z., P.J., and J.H. performed XPS measurements and wrote corresponding part of the work. L.D. performed TEM and EDS studies. O.V.L., V.N.P. and I.V.P. sintered the $(Hf,Zr)O_2$ nanopowders. A.O.D., M.D.V., O.S.P. and M.P.T. performed the dielectric measurements. M.V.K. performed XRD studies. E.A.E. and A.N.M. performed theoretical modelling and the draft of the work. Corresponding authors made all improvements in the manuscript.

## Acknowledgments

The work of Y.O.Z., E.A.E., A.O.D. and M.D.V. is funded by the National Research Foundation of Ukraine (grant N 2023.03/0127 "Silicon-compatible ferroelectric nanocomposites for electronics and sensors"). A.N.M. and O.S.P. acknowledge the support the National Research Foundation of Ukraine (grant N 2023.03/0132 "Manyfold-degenerated metastable states of spontaneous polarization in nanoferroics: theory, experiment and perspectives for digital nanoelectronics"). P.J. and J.H. acknowledge the support of the Strategy AV21 project: "Study of the atomically thin quantum materials by advanced microscopic/spectroscopic techniques applying machine learning". L.D. acknowledges support from the Knut and Alice Wallenberg Foundation (grant no. 2018.0237) for TEM research. Results of theoretical modelling were visualized in Mathematica 15.0 [35].